\documentclass[twoside,11pt]{article}
\usepackage[english]{babel}
\usepackage{latexsym}
\usepackage{makeidx}
\usepackage{amssymb}

\def\Ro{{\mathbb R}}

\newcommand{\eqnsection}{
\renewcommand{\theequation}{{\thesection.\arabic{equation}}}
\makeatletter
\csname @addtoreset\endcsname{equation}{section}
\makeatother}
\eqnsection
\newcommand{\be}{\begin{equation}}
\newcommand{\en}{\end{equation}}
\newcommand{\beg}{\begin{eqnarray}}
\newcommand{\ene}{\end{eqnarray}}
\newcommand{\ee}{\end{equation}}
\newcommand{\irr}[1]{\hspace{0.2em}
  \stackrel{
   \setbox0=\hbox{\hspace{0.06em}$\displaystyle #1$\hspace{0.06em}}
   \setbox1=\hbox{\vrule width\wd0 height0.08ex depth0pt}
   \vrule width0.08ex height0.08ex depth0.475ex \box1
   \vrule width0.08ex height0.08ex depth0.475ex }{#1}
  \hspace{0.2em}}

\newcommand{\mvec}[1]{\mbox{\boldmath{$#1$}}}
            
            \newcommand{\m}{\mvec{m}}
\newcommand{\bee}[1]{\begin{equation}\label{#1}}
            \newcommand{\bey}{\begin{eqnarray}}
            \newcommand{\byy}[1]{\begin{eqnarray}\label{#1}}
            \newcommand{\eey}{\end{eqnarray}}
            \newcommand{\beo}{\begin{eqnarray*}}
            \newcommand{\eeo}{\end{eqnarray*}}

            \newcommand{\argm}{(\m ,\mvec{x}, t)}
            
\newcommand{\X}{\!\mvec{X} (\cdot)}

\title{A mesoscopic approach  to diffusion phenomena in mixtures}
\author{A.Palumbo$^1$, C. Papenfuss$^2$  and P. Rogolino$^1$\\
$^1$Department of Mathematics, University of Messina, Italy\\
$^2$ Technische Universit\"at Berlin
Institut f\"ur Mechanik\\ Stra\ss e des 17. Juni 135 \\D-10623
Berlin, Germany}

\date{}
\begin{document}
\maketitle
\begin{abstract}
The mesosocpic concept is applied to the theory of mixtures. The
aim is to investigate the diffusion phenomenon from a mesoscopic
point of view. The domain of the field quantities is extended by
the set of mesoscopic variables, here the velocities of the
components. Balance equations on this enlarged space are the
equations of motion for the mesoscopic fields.  Moreover,  local
distribution functions of the velocities are introduced as a
statistical element, and an equation of motion for this
distribution function is derived. From this equation of motion
differential equations for the diffusion fluxes, and also for
higher order fluxes are obtained. These equations are of balance
type, as it is postulated in Extended Thermodynamics. The
resulting  evolution equation for the diffusion flux  generalizes
the Fick's law.
\end{abstract}
\section{Introduction}
Most phenomena involving mass diffusion are described by means of
Fick's law which relates linearly the mass flow to the gradient of
the mass concentration through the diffusion coefficient. In
practical applications Fick's law is satisfactory to a large
extent. However, it is not applicable to transitory situations
involving for instance high frequencies. The mass balance together
with Ficks law leads to parabolic differential equations for the
concentrations, allowing for  infinite velocities of propagation
of disturbances. The last paradox must be overcome by considering
non-Fickian transport behavior for the diffusive mass flux. A
possible solution is given in Extended Thermodynamics
 \cite{Jou,Jou92,Jou96,15Jou,MuellerRuggeri}. The diffusion flux is
introduced as an additional field variable,  obeying  a differential equation. This differential equation is derived either
 from an Irreversible Thermodynamics treatment \cite{Lebon01, Garcia_94}
  of the dissipation inequality (for other applications
  see \cite{Garcia_93, Jou_1,Lebon_94,Casas_93,Criado_92,Criado_93}), or a balance type
  equation for the non-convective fluxes, here the diffusion flux, is postulated in Rational Extended Thermodynamics
  \cite{MULLER,MuellerRuggeri,Mueller_Rev}. For a purely macroscopic derivation of balance type differential equations for
  higher order fluxes see \cite{Van_Cimelli_04,Van_Cimelli_05}. Within
  Rational Extended Thermodynamics it has been shown that the resulting system of field equations is symmetric hyperbolic
   with a convex extension \cite{Ruggeri83,MuellerRuggeri,Weiss_dr}, thereby allowing for finite speeds of disturbances
only \cite{Friedrichs_Lax,Lax73,Ruggeri83}.

 A foundation of the field equations of Extended Thermodynamics has been given from statistical mechanics. In most cases
 the single particle distribution function is used, together with Boltzmann-equation
 \cite{MuellerRuggeri}. In this case
 microscopic interactions  are introduced via the collision term in the Boltzmann-equation. However, Boltzmann-equation is,
 strictly speaking, valid only for dilute gases, and for  liquids or solids a different foundation is necessary. Such a
 background theory can be given using the BBGKY-hierarchy
 (\cite{BornGreen46,BornGreen47,Bogoljuboff,Kirkwood}) for the
 hierarchy of N-particle distribution functions. The structure of the field equations of Extended Thermodynamics could be
 derived from this statistical mechanics background. Such an approach necessitates the knowledge of microscopic inter-particle
 interaction potentials.
      \newline

An approach, different from microscopic statistical
mechanics, is the so called mesoscopic theory
\cite{Sitges,MUPAEH96,poz}.
It is a continuum theory, introducing constitutive functions on a continuum level, and no microscopic inter-particle
interactions. On the other hand the mesoscopic
description is finer than the macroscopic description. This additional
information leads to the definition of internal variables
in complex media, and to equations of motion for them. This
mesoscopic concept has been applied to
liquid crystals
\cite{BLENK91,BLENK92,PHYSICA,PhysRev95,Journal,Budapest2,
MUPAEH00},
to damage and fracture mechanics \cite{mesocrack,VanPa02},
and to dipolar media \cite{PACIRO02}.
We will show in the following that within the mesoscopic
theory a balance type equation for the diffusion flux can be
 derived.
\newline
The basic idea of the mesoscopic theory is to introduce an
additional variable, the so-called mesoscopic variable, in the
domain  of the field quantities. On this higher dimensional space
a continuum theory is developed. In this theory the mesoscopic
field quantities depend not only on position and time, but also on
the  mesoscopic variable. In addition a local distribution
function of the mesoscopic variable is introduced  as a
statistical element. Macroscopic fields are the averages of
mesoscopic ones over the additional (mesoscopic) variable.

The aim of this paper is the application of the mesoscopic concept
to an N-component mixture and a derivation of differential
equations for the diffusion fluxes from the mesoscopic theory. The
application of the resulting equations to fast phenomena,
involving diffusion processes is left for future work, as well as
the investigation whether our generalized diffusion equation leads
to hyperbolic field equations.

\section{Mesoscopic concept and balance equations}
The mesoscopic concept introduces the mesoscopic space
$$(\mvec{ m},\mathbf x,t)\in \mathcal M\times\Ro^3_x\times \Ro_t$$
on which all field quantities  are defined. Here $\mvec{ m}$ is a
set of mesoscopic variables which is an element of a suitable
manifold $\mathcal M$ on which an integration can be defined.
Beyond the use of additional variables $\mathbf m$ the mesoscopic
concept introduces a statistical element, the so-called {\it
mesoscopic distribution function} (MDF) $f(\mathbf m,\mathbf x,t)$
generated by the different values of the mesoscopic variable of
the particles in a volume element
\begin{equation}
f(\mathbf m,\mathbf x,t)\equiv f(\cdot),\,\,\,\,(\cdot)\equiv
(\mathbf m,\mathbf x,t)\in \mathcal M\times R^3_x\times R_t.
\end{equation}
The MDF describes the distribution of $\mathbf m$ in a volume
element around $\mathbf x$ at time $t$, and therefore it is
normalized
\begin{equation}\label{eq:1}
\int f(\mathbf m,\mathbf x,t) d\mathbf m =1.
\end{equation}
It is the probability density of having the particular value $\mathbf m$
of the mesoscopic variable in the continuum element at position $\mathbf x$
and time $t$. Therefore it is the fraction
\begin{equation}
f(\mathbf m,\mathbf x,t)=\frac{\rho(\mathbf m,\mathbf
x,t)}{\rho(\mathbf x,t)}.
\end{equation}
Here $\rho(\mathbf x,t)$ is the macroscopic mass density. By use of
(\ref{eq:1}) we obtain
\begin{equation}
\rho(\mathbf x,t)=\int \rho(\mathbf m,\mathbf x,t) d\mathbf m.
\end{equation}

\subsection{Balance equations}

Let $G$ denote  a region in $\mathbb R^3 \times \mathcal{ M} $ and
$\mvec{X}$ the density of an extensive quantity. Then the  global
quantity in the region $G$ changes due to a flux over the boundary
of $G$ and due to production and supply within $G$: \be
\frac{d}{dt} \int_G \mvec{X} d^3 x d\mvec{m} = \int_{\partial G}
\mvec{\phi_X}(\cdot )da + \int_G \Sigma_X (\cdot ) d^3x d\mvec{m}
\quad .\ee A generalized Reynolds transport theorem in the
mesoscopic  space, analogous to the one in  \cite{Ehrentraut_diss}
is used to transform the time derivative, and Gauss theorem is
applied to the boundary integral. The boundary $\partial G$  of
$G$ consists of a boundary in position space and a boundary on the
manifold $\mathcal{ M} $. Then we have  in regular points of the
continuum the {\em local mesoscopic balance} \cite{MCLC}
\bee{B:20} \frac{\partial}{\partial t}\X + \nabla_x \cdot
[\mathbf{v}(\cdot )\X - \mvec{S}(\cdot)] + \nabla_m \cdot
[\mvec{w}(\cdot)\X - \mvec{R}(\cdot)] = \mvec{\Sigma}(\cdot). \ee
Here the independent field $\mvec{w}(\cdot)$, defined on the
mesoscopic space, describes the change in time of the set of
mesoscopic variables: With respect to $\m$ the {\em mesoscopic
change velocity} $\mvec{w}(\cdot)$ is the analogue to the
mesoscopic material velocity $\mathbf{v}(\cdot )$ referring to
$\mvec{x}$: If a particle is characterized by $\argm$, then for
$\Delta t\rightarrow + 0$ it is characterized by $(\mvec{m} +
\mvec{w}(\cdot)\Delta t, \mvec{x} + \mathbf{v}(\cdot )\Delta t, t
+ \Delta t)$ at later times $t + \Delta t$.  Besides the usual
gradient, also the gradient with respect to the set of mesoscopic
variables appears. This expresses the fact that there is a flux
over the boundary on the manifold $\mathcal M$.\newline According
to equation (\ref{eq:1}) we obtain from the mesoscopic mass
balance a balance type differential equation for  the MDF
$f(\cdot)$ by inserting its definition \cite{Bam,poz}:
\begin{eqnarray}
&&\frac{\partial}{\partial t} f(\cdot) + \nabla_{x}\cdot
[\mathbf{v} (\cdot) f(\cdot)] +
\nabla_{m}\cdot [\mbox{\boldmath{$w$}} (\cdot) f(\cdot)] + \nonumber\\
&&+ f(\cdot) \left [\frac{\partial}{\partial t} +
\mathbf{v}(\cdot) \cdot
  \nabla_{x}\right ] \ln \varrho(\mbox{\boldmath{$x$}},t) =0 .\label{B:21}
\end{eqnarray}

\section{Application of the mesoscopic concept to mixtures}

Let us consider a mixture of N different chemical components. The
different components are distinguished by a component index in
capital letters. We will introduce velocity distributions of all
chemical components, denoted $f^A$. Each of them is a function of
position of the respective continuum element, time, and particle
velocity. $f^A(\mathbf x, t, \mathbf v)$ gives the probability
density of finding the value $\mathbf{v}$ for the material
velocity  of component A of the mixture in a volume element around
$\mathbf x$ at time $t$.



 The mesoscopic mass
densities and all other mesoscopic fields are different for
different chemical components. Mesoscopic fields are denoted by a
"hat", where this is necessary in order to distinguish them from
the corresponding macroscopic quantities.\newline The distribution
function $f^A$ is defined as follows:
\begin{equation}\label{eq:fdA}
f^A(\mathbf x,t,\mathbf v)=:\frac{\hat\rho^A(\mathbf x,t,\mathbf v
)} {\rho^A(\mathbf x,t)}\ .
\end{equation}
$f^A(\mathbf x , t,\mathbf v) d\mathbf v$ gives the fraction of
particles of component $A$ having a material velocity in the
element $d\mathbf v$ on $\Ro^3_v$ around $\mathbf v$.

The distribution functions are normalized:
\begin{equation}
\int_{\Ro^{3}_v} f^A(\mathbf x, t,\mathbf v) d\mathbf
v=1,\,\,\,\,\:\forall A \ .
\end{equation}

\section{Macroscopic balance equations for the  mixture and its components}
In this section,  we shall recall briefly the   macroscopic
balance equations for mixtures and for the different
components.\newline The total mass of a mixture is a conserved
quantity, while the mass of a particular component is not
conserved, if chemical reactions occur. The same happens for
momentum, and energy due to chemical reactions between different
components of the mixture. Therefore, in the balance equations for
components there are production terms on the right hand side.

\subsection{Macroscopic balance equations for the mixture}

The balance equations for the mixture as a whole look the same as
the balance equations for a one-component system
\cite{MUSCHIK83a1,MUSCHIK83a2}. From the structure of balance
laws, mixtures cannot be distinguished from chemically
 pure substances.

{\bf Balance of
mass}\newline
\begin{equation}
\frac{\partial \rho}{\partial t}(\mathbf x,t)+\nabla\cdot(\rho
\mathbf{v})(\mathbf x,t)=0
\end{equation}
Here $\mathbf{v}$ is the macroscopic  material  velocity of the
mixture, i.e., the barycentric velocity of all particles in the
volume element.
\newline {\bf Balance of momentum}\newline \be\label{eq:mom}
\frac{\partial}{\partial t}(\rho\mathbf{v})+\nabla \cdot (\rho
\mathbf{v}\otimes\mathbf{v} -\mathbf t^{^T}) =\rho \mathbf f\ ,
\ee where $\mathbf t$ denotes Cauchy stress tensor, and $\mathbf
f$ is the acceleration (specific density of volume forces).
{\bf
Balance of energy}\newline \beg\nonumber
&&\frac{\partial}{\partial t}[\rho(e +\frac 1{2}\mathbf{v}
\cdot\mathbf{v})]+\nabla\cdot [\mathbf q-\mathbf{v}\cdot \mathbf
t+\rho\mathbf{v}(e +\frac 1{2}\mathbf{v}\cdot\mathbf{v})]
\\ &&-\rho(\mathbf f\cdot \mathbf{v}+ r)=0  \label{eq:en} \ene where  $e$ is
the specific internal energy density (internal energy per unit
mass), $\mathbf q$ denotes the heat flux density, and $r$ is the
energy supply density.

\subsection{Macroscopic balance equations for the components}
{\bf Balance of mass of component $A$}\newline The local balance
of mass of component $A$, taking into account production of mass
of that component due to chemical reactions, reads:
\begin{equation}       \label{eq:mass}
\frac{\partial \rho^A}{\partial t}(\mathbf x,t)+\nabla\cdot(\rho^A
\mathbf{v}^A)(\mathbf x,t)=P^A_{chem}
\end{equation}
here $\mathbf{v}^A$ is the material velocity of component $A$. The
mass densities of the different chemical components are additive:
\be \sum_A\rho^A=\rho \ee The material velocity $\mathbf{v}$ of
the mixture is defined as the weighted sum of the component
velocities:
\begin{equation}\label{eq:frac}
\sum_A\rho^A\mathbf{v}^A=\rho\mathbf{v} \label{sum_vA}
\end{equation}
Equation (\ref{eq:mass}) may be cast in a more useful form in
terms of mass fractions $c^A$ defined by $c^A =
\frac{\rho^A}{\rho}$, $A=1,..., N$ and the diffusion fluxes
$\mathbf J^A$
\begin{equation}\label{eq:flux}
\mathbf J^A=\rho^A(\mathbf{v}^A-\mathbf{v})
\end{equation}
such that $\sum_A\mathbf J^A=0$.
The mass fraction balance equation is
\begin{equation}
\rho\frac{d}{dt} c^A+\nabla\cdot \mathbf J^A=P^A_{chem}
\end{equation}
where $\frac d{dt}=\frac{\partial}{\partial
t}+\mathbf{v}\cdot\nabla$ is the material time derivative.\newline
{\bf Balance of momentum of component $A$}\newline The momentum
balance for one particular chemical component is: \be
\label{eq:momA} \frac{\partial}{\partial
t}(\rho^A\mathbf{v}^A)+\nabla \cdot (\rho^A
\mathbf{v}^A\otimes\mathbf{v}^A -\mathbf t^{A{^T}}) =\rho^A
\mathbf f^A +\mathbf P^A_m. \ee with a production $P^A_m$ of
momentum of component $A$.
\newline
{\bf Balance of energy of component $A$}\newline The energy
balance equation is given by: \beg\nonumber
&&\frac{\partial}{\partial t}[\rho^A(e^A +\frac 1{2}\mathbf{v}^A
\cdot\mathbf{v}^A)]+\nabla\cdot [\mathbf q^A-\mathbf{v}^A\cdot
\mathbf t^A+\rho^A\mathbf{v}^A(e^A +\frac 1{2}\mathbf{v}^A\cdot\mathbf{v}^A)]\\
&&-\rho^A(\mathbf f^A\cdot \mathbf{v}^A+ r^A) =P_e^A \ene with a
component energy production due to chemical reactions.  In this
set of balance equations constitutive equations for the component
stress tensor, component internal energy density, and heat flux
density, and the productions due to chemical reactions are needed
in order to close the system of differential equations. These
constitutive quantities must be related to the wanted fields in a
material dependent manner. The
 constitutive theory, including restrictions on constitutive functions from general principles, is out of the scope of the
 present paper.

The (total) energy density of the mixture is the sum over the
components. From the additivity of the extensive quantities mass,
momentum, and energy  follows that the balances of the mixture are
obtained summing up the component equations
\cite{MUSCHIK83a1,MUSCHIK83a2}. This leads to relations between
constitutive quantities for components and constitutive quantities
for the mixture.

\section{Mesoscopic balance equations for the mixture and its components}

Mesoscopic balance equations are defined on the mesoscopic space.
For the mixture as a whole the additional mesoscopic variable is
the material velocity $\mathbf{v}$ of the mixture, which is the
weighted sum of the component velocities, see equation
(\ref{sum_vA}). The mesoscopic balance equations are derived
analogously to the derivation discussed in the second section (see
 equation (\ref{B:20}) for the general structure of these balances) and by suitable identifications. The mesoscopic balance
 equations for the mixture are those derived previously for one component systems (see Section 2). In all cases it is
 supposed that the constituents of the mixture have no internal angular momentum, i.e., we are not dealing with micropolar
continua.

For the mixture we introduce the distribution function of the
material velocity: \be f(\mvec{x},t,\mathbf{v})= f(\bullet ) =
\frac{\hat\rho(\bullet )}{\rho (\mvec{x},t)} \label{eq:fd}\en

Explicitly we have  the following mesoscopic balances:\newline
\subsection{Mesoscopic balance equations for the mixture}
{\bf Mesoscopic balance of mass}\newline
\begin{equation}\label{eq:4}
\frac{\partial}{\partial t}(\hat\rho)(\bullet )
+\nabla\cdot(\hat\rho\mathbf v)(\bullet ) +\nabla_{\mathbf
v}\cdot(\hat \mathbf w\hat\rho)(\bullet )=0
\end{equation}
here $(\bullet )$ is the abbreviation for
$(\mvec{x},t,\mathbf{v})$, $\mathbf v$ and $\mathbf w $ are the
mesoscopic velocity of the mixture and the change in time of the
mesoscopic variable $\mathbf v$, respectively ($ \mathbf{w} =
\mathbf{ \dot{v}} $).\newline According to the definition of the
distribution function  equation (\ref{eq:fd}), we obtain from the
mesoscopic mass balance (\ref{eq:4}) a differential equation for
the MDF $f(\bullet)$: \beg\label{eq:funcdis1} &&\frac{\partial
f(\bullet)}{\partial t}+\nabla\cdot(f(\bullet)\mathbf v)
+\nabla_{\mathbf v}(f(\bullet)\mathbf w(\bullet))+\\\nonumber
&&+f(\bullet)\Big(\frac{\partial}{\partial t}+\mathbf v\cdot
\nabla\Big) \lg\rho(\mathbf x,t)=0 \ene {\bf Mesoscopic balance of
momentum} \beg\label{eq:7} \frac{\partial}{\partial
t}(\hat\rho\mathbf v)+\nabla\cdot (\hat\rho \mathbf v\otimes
\mathbf v)+\nabla_{\mathbf v}\cdot (\hat\rho \mathbf{ \hat
w}\otimes \mathbf v)-\nabla\cdot\mvec{\hat t}^T-\nabla_{\mathbf v}
\cdot\hat {\mathbf T}^T= \hat\rho\hat\mathbf f. \ene with
$\mvec{\hat t}^T= \mvec{\hat t}$
 and $\mvec{\hat T}^T=
\mvec{\hat T}$. Here $\hat\mathbf T$ is the analogue  in velocity
space of the stress tensor, i.e., the non-convective momentum flux
in velocity space. All field quantities here depend on position,
time and particle velocity.

 Due to definition (\ref{eq:fd}) the macroscopic material velocity
of the mixture, $\mathbf{v}(\mathbf x,t)$, i.e., the mean velocity
of the mixture, is expressed as follows: \be\label{eq:mesvel}
\rho(\mathbf x,t)\mathbf{v}(\mathbf x,t)= \int_{\Ro^3} \hat\rho
\mathbf v d\mathbf v , \ee {\bf Mesoscopic balance of energy}
\beg\nonumber &&\frac{\partial}{\partial t}[\hat\rho(\hat e +\frac
1{2}\mathbf v \cdot\mathbf v)]+\nabla\cdot [\hat\mathbf q-\mathbf
v\cdot\hat \mathbf t+\rho(\hat e +\frac 1{2}\mathbf v\cdot\mathbf
v) \mathbf v]+ \\\nonumber &&+\nabla_{\mathbf v}\cdot[\hat\mathbf
Q-\mathbf w\cdot \hat\mathbf T+\hat\rho\mathbf w(\hat e+\frac{
1}{2}\mathbf v\cdot \mathbf v)]=\\       \label{eq:enmes}
&&=\hat\rho(\hat\mathbf f\cdot \mathbf v+ \hat r) +\hat P_e \ene
$\mathbf Q$ is the analogue to the heat flux density in velocity
space, i.e., a non-convective energy flux over the boundary in the
mesoscopic part of the domain, which consists of position, time
and velocity for all fields, here.

By integrating these mesoscopic balances over the mesoscopic
space, i.e., over all velocities, with the meosocopic distribution
function as statistical weight,  the macroscopic balance equations
for the mixture are obtained.

\subsection{Mesoscopic balance equations  for the components of the mixture}
{\bf Mesoscopic balance of mass of component A} \newline By using
the general form of the mesoscopic balance of mass (see also
\cite{PHYSICA,Blenk93,Bam,poz}), we have the following equation
for the component A:
\begin{equation}\label{eq:2}
\frac{\partial}{\partial t}(\hat\rho^A)(\cdot )
+\nabla\cdot(\hat\rho^A\mathbf v^A)(\cdot ) +\nabla_{\mathbf
v}\cdot(\mathbf w^A\hat\rho^A)(\cdot )=\hat P^A_{chem}(\cdot )
\end{equation}
here $\mathbf v^A$ represents the material velocity of component
$A$,  and $\mathbf{ w^A}$ is the change in time of the mesoscopic
variable $\mathbf v^A$: $ \mathbf{ w^A} = \mathbf{\dot v^A}$. This
acceleration $\mathbf{ w^A}$ is a mesoscopic constitutive
function. The term on the right hand side represents a production
due to chemical reactions.

The sum of the mass balances of the different components is the
mass balance of the mixture, because the mass densities are
additive:
\begin{equation}\label{eq:3}
\sum_{A=1,...,N}\hat\rho^A=\hat\rho.
\end{equation}
\begin{equation}\label{sum}
\frac{\partial}{\partial
t}\sum_A(\hat\rho^A)+\nabla\cdot\sum_A(\hat\rho^A\mathbf v^A)
+\nabla_{\mathbf v}\cdot\sum_A(\mathbf w^A\hat\rho^A)=\sum_A\hat
P^A_{chem} \ . \end{equation}

By comparing in equations (\ref{sum}) and (\ref{eq:4}) the fluxes
in mesoscopic space and in position space we obtain the following
relations: \beg\label{eq:5} \sum_A \nabla_{\mathbf v}\cdot(\mathbf
w^A\hat\rho^A)=\nabla_{\mathbf v}\cdot (\mathbf w\hat\rho)\ ,
\\\label{eq:51} \sum_A \nabla\cdot(\hat\rho^A\mathbf v^A)=\nabla
\cdot (\hat\rho\mathbf v)\ , \ene and for the production terms:
\be \sum_A\hat P^A_{chem} =0\ .\ee A solution of  equations
(\ref{eq:5}) and (\ref{eq:51}) is: \beg
\mathbf w=\sum_A\frac{\hat\rho^A}{\hat\rho}\mathbf w^A,\\
\mathbf v=\sum_A\frac{\hat\rho^A}{\hat\rho}\mathbf v^A
\ene

{\bf Mesoscopic distribution function for the  components}\newline
The distribution function $f^A(\mathbf x,t,\mathbf v )$ gives the
probability density of finding a particle of component $A$ with
velocity $\mathbf v^A$ in the volume element around $\mathbf x$ at
time $t$. It is the mass fraction \be f^A(\mathbf x,t,\mathbf v )=
\frac{\hat \rho^A (\mathbf x,t,\mathbf v)}{\rho^A (\mathbf
x,t)}\quad .\en The mesoscopic balance of mass for component $A$,
equation (\ref{eq:2}), can be used to derive an equation of motion
for the distribution function: \beg\label{eq:funcdis}
&&\frac{\partial f^A}{\partial t}+\nabla\cdot(f^A\mathbf v^A)
+\nabla_{\mathbf v}\cdot (f^A\mathbf w^A)+\\\nonumber
&&+f^A\Big(\frac{\partial}{\partial t}+\mathbf v^A\cdot
\nabla\Big) \lg\rho^A(\mathbf x,t) =\frac{\hat
P^A_{chem}}{\rho^A(\mathbf x,t)} \ene Introducing the macroscopic
fields by integrating over the mesoscopic part  of the enlarged
domain, here the velocity, with the component distribution
function as probability density,
 we can transform local mesoscopic balances into local macroscopic balances.
So, the definition of the macroscopic material velocity of the
component $A$, $\mathbf{v}^A(\mathbf x,t)$, i.e., the mean
velocity of the component $A$ of the mixture, is
\begin{equation}                            \label{eq:mesvelA}
\rho^A\mathbf{v}^A(\mathbf x,t)= \int_{\Ro^3} \hat\rho^A \mathbf v
d\mathcal M,
\end{equation}
and equation (\ref{eq:2}) leads to the macroscopic mass balance of
component $A$ (\ref{eq:mass}) if we take
\begin{equation}
\int\nabla_{\mathbf v}\cdot [\mathbf w^A\hat\rho^A] d\mathcal M=0
\end{equation}
into consideration. This holds because of Gauss' theorem and
because $f^A(\cdot)$, restricted to the mesoscopic part has a
compact support.
\newline
\vspace*{0.5cm}

 {\bf Relation between the mesoscopic distribution
functions of the components and the distribution function of the
mixture} \vspace*{0.3cm}

The velocity distribution function of a particular component $f^A$
was defined as \be f^A(\mathbf x, t, \mathbf v)= \frac{\hat
\rho^A(\mathbf x, t, \mathbf v)}{\rho^A(\mathbf x, t)} \ , \en and
the velocity distribution of the mixture is \be f(\mathbf x, t,
\mathbf v)= \frac{\hat \rho(\mathbf x, t, \mathbf v)}{\rho(\mathbf
x, t)} \ . \en

Because the mass densities of the different chemical components
are additive: \be \sum_A \hat \rho^A(\mathbf x, t, \mathbf v)=\hat
\rho (\mathbf x, t, \mathbf v)\ , \quad \sum_A \rho^A(\mathbf x,
t)= \rho (\mathbf x, t)\ ,\en we end up with a relation between
the distribution functions of the components and that of the
mixture: \be \sum_A f^A(\mathbf x, t, \mathbf v)\rho^A (\mathbf x,
t) = \sum_A \hat \rho^A(\mathbf x, t, \mathbf v)= \hat \rho
(\mathbf x, t, \mathbf v)=f(\mathbf x, t, \mathbf v) \rho (\mathbf
x, t) \ ,\en or \be  \sum_A f^A(\mathbf x, t, \mathbf
v)\frac{\rho^A (\mathbf x, t)}{\rho (\mathbf x, t)} =f(\mathbf x,
t, \mathbf v)\ . \en The distribution function of the mixture is
the sum of the component distribution functions, weighted with the
mass fractions of the components.

{\bf Mesoscopic balance of momentum of
component A}\newline \be          \label{eq:6}
\frac{\partial}{\partial t}(\hat\rho^A\mathbf v^A)+\nabla \cdot
(\hat\rho^A \mathbf v^A\otimes\mathbf v^A) +\nabla_{\mathbf
v}\cdot (\hat\rho^A\mathbf w^A\otimes \mathbf
v^A)-\nabla\cdot\hat\mathbf t^A-\nabla_{\mathbf v}\cdot \hat
\mathbf T^A=\hat\rho^A \hat\mathbf F^A +\hat\mathbf P^A_m. \ee The
mesoscopic momenta $\hat\rho^A\mathbf v^A$ are additive, and
summing up equations (\ref{eq:6}) over the different components
one obtains the  mesoscopic balance of momentum (\ref{eq:7}). By
comparing the fluxes in position space and in velocity space,
respectively, we have: \beg           \label{eq:8} \hat\mathbf
t=\sum_A(\hat\mathbf t^A-\hat\rho^A\delta\mathbf v^A\otimes
\delta\mathbf v^A ),\\\label{eq:9} \hat\mathbf
T=\sum_A(\hat\mathbf T^A-\hat\rho^A\delta\mathbf w^A\otimes
\delta\mathbf v^A ), \ene where we have  introduced the
abbreviations \be \delta\mathbf v^A=\mathbf v^A-\mathbf
v,\,\,\,\,\: \delta\mathbf w^A=\mathbf w^A-\mathbf w. \ee Finally,
a comparison of the production terms in the mesoscopic balances of
momentum for the sum of components on one hand, and for the
mixture on the other hand,  leads to the relation: \be
\hat\rho\hat\mathbf f=\sum_A(\hat\rho^A\hat\mathbf f^A
+\hat\mathbf P^A_m) \ee By integrating over the mesoscopic space
the macroscopic momentum balance for component $A$,  equation
(\ref{eq:momA}) is  obtained.\newline {\bf Mesoscopic balance of
energy of component A}\newline For particles of component $A$ and
particle velocity $\mathbf v^A$ we have: \beg \nonumber
&&\frac{\partial}{\partial t}[\hat\rho^A(\hat e^A +\frac
1{2}\mathbf v^A \cdot\mathbf v^A)]+\nabla\cdot [\hat\mathbf
q^A-\mathbf v^A\cdot\hat \mathbf t^A+\rho^A(\hat e^A +\frac
1{2}\mathbf v^A\cdot\mathbf v^A) \mathbf v^A]+\\   \nonumber &&+
\nabla_{\mathbf v^A}\cdot[\hat\mathbf Q^A-\mathbf w^A\cdot
\hat\mathbf T^A+\hat\rho^A\mathbf w^A(\hat e^A+\frac 1{2}\mathbf
v^A\cdot
\mathbf v^A)]=\\
&&=\hat\rho^A(\hat\mathbf f^A\cdot \mathbf v^A+ \hat r^A) +\hat
P_e^A \ene As before, summing over all components of the mixture
one obtains the mesoscopic balance of energy (\ref{eq:enmes}) and
the integration over $\Ro^3$ leads to the macroscopic balance of
energy expressed by equation (\ref{eq:en}).

\section{Constitutive quantities for the components compared to those for the mixture}

Analogously to the previous section  the flux terms and the
production terms in the balance equations of the mesoscopic
mixture and the sum of the component equations are compared. As a
result the internal energy density of the mixture, the heat flux
and its analogue in mesoscopic space, and the energy absorption
density of the mixture are expressed by the following relations:
\beg
&&\hat\rho\hat e=\sum_A(\hat\rho^A\hat e^A +\frac 1{2}\hat\rho^A\delta\mathbf v^A\cdot\delta\mathbf v^A),\label{6.1}\\
&&\hat\mathbf q=\sum_A\{\hat\mathbf q^A-\delta\mathbf v^A\cdot\hat\mathbf
t^A +\hat\rho^A\hat e^A\delta\mathbf v^A+\frac 1{2}[\hat\rho^A(\delta
\mathbf v^A\cdot\delta\mathbf v^A)\delta\mathbf v^A]\},\\\nonumber
&&\hat\mathbf Q=\sum_A\{\hat\mathbf Q^A-\hat\rho^A(\mathbf w\cdot \delta
\mathbf w^A)\delta\mathbf v^A-\delta\mathbf w^A\cdot \hat\mathbf T^A+
\hat\rho^A\hat e^A\delta\mathbf w^A+ \\
&&+\hat\rho^A(\delta\mathbf v^A\cdot
\mathbf v)\delta\mathbf w^A+\frac 1{2}[\hat\rho^A(\delta\mathbf v^A
\cdot \delta\mathbf v^A)\delta\mathbf w^A]\},\\
&&\hat\rho\hat r=\sum_A(\hat\rho^A\hat r^A+\hat\rho^A\mathbf v^A\cdot
\delta\hat\mathbf f^A) \quad \mbox{with} \quad \delta\hat\mathbf f^A = \hat\mathbf f^A- \hat\mathbf f ,\\
&&\hat P_e=\sum_A \hat P^A_e.\label{6.5} \ene Equations
(\ref{6.1}) to (\ref{6.5}) show that  the mesoscopic constitutive
quantities of the mixture are not, in general, the sum of the
corresponding constitutive quantities of the components, but some
fluctuation terms contribute.

\section{Evolution equations for diffusion fluxes}
The aim of this section is to obtain evolution equations for
diffusion fluxes of the different components.
By using relations (\ref{eq:mesvel}), (\ref{eq:mesvelA})
and the definitions (\ref{eq:fdA}), (\ref{eq:fd})
it is possible to
define the diffusion flux $\mathbf J^A$, expressed by equation
(\ref{eq:flux}), as follows:
\be
\mathbf J^A(\mathbf x,t)=\rho^A(\mathbf x,t)\int_{\Ro^3}[f^A(\cdot)-
f(\cdot)]\mathbf v d\mathcal M
\ee
In order to derive evolution equations for fluxes let us
introduce the family of the macroscopic
fields of order parameters  which is defined by different moments of the
distribution functions  $f^A(\cdot)$ and $f(\cdot)$.
They are defined as
\beg
&&\mathbf a^A_k=\int_{\Ro^3}f^A(\cdot)\underbrace{\irr
{\mathbf v\dots\mathbf v}}_k d\mathcal M,\\
&&\mathbf a_k=\int_{\Ro^3}f(\cdot)\underbrace{\irr {\mathbf
v\dots\mathbf v}}_k d\mathcal M, \ene where $\irr{\dots}$ denotes
the symmetric irreducible (traceless) part of a tensor
\cite{PHYSICA}. These fields of order parameters describe
macroscopically the mesoscopic state of the system introduced by
$\mathbf v^A$ and $\mathbf v$ and its distribution functions $f^A$
and $f$. Thus these are the link between the mesoscopic background
description of the system and its extended description by
additional macroscopic fields. In a macroscopic phenomenological
theory they represent in general internal variables which satisfy
relaxation equations. In our case the first moments $\mathbf a_1$
and $\mathbf a_1^A$ are not internal variables, but classical
wanted fields, namely the specific momentum densities (material
velocities) $\mathbf{v}$ and $\mathbf{v}^A$. The higher order
moments are internal variables in the sense of thermodynamics.

 The diffusion
fluxes $\mathbf J^A,\,\,\,(A=1,...,N)$ are proportional to the
difference of the first moment of the distribution functions
$f^A(\cdot)$ and $f(\cdot)$. If we multiply equations
(\ref{eq:funcdis}) and (\ref{eq:funcdis1}) with the mesoscopic
variable  and integrate over the manifold $\mathcal M$ we obtain,
respectively: \beg\nonumber &&\int_{\Ro^3}\frac{\partial}{\partial
t} [f^A(\cdot)\mathbf v^A] d\mathcal M+
\int_{\Ro^3}\nabla\cdot[\mathbf v^A f^A(\cdot)\mathbf v^A]
d\mathcal M+\\\nonumber &&+\int_{\Ro^3}\mathbf v^A\nabla_{\mathbf
v}\cdot[\mathbf w^A(\cdot) f^A(\cdot)]d\mathcal M+
\int_{\Ro^3}f^A\mathbf v^A\Big[\frac{\partial}{\partial t}
+\mathbf v\cdot\nabla\Big] \lg\rho^A(\mathbf x,t)d\mathcal M=\\
&&= \int_{\Ro^3}\frac{\hat P^A_{chem}}{\rho^A(\mathbf x,t)}
d\mathcal M \ene and \beg\nonumber
&&\int_{\Ro^3}\frac{\partial}{\partial t} [f(\cdot)\mathbf v]
d\mathcal M+ \int_{\Ro^3}\nabla\cdot[\mathbf v f(\cdot)\mathbf v]
d\mathcal M+\\\nonumber &&+\int_{\Ro^3}\mathbf v\nabla_{\mathbf
v}\cdot [\mathbf w(\cdot) f(\cdot)]d\mathcal M+ \int_{\Ro^3}f
\mathbf v\Big[\frac{\partial}{\partial t} +\mathbf
v\cdot\nabla\Big] \lg\rho(\mathbf x,t)d\mathcal M=0. \ene The
mesoscopic manifold $\mathcal M$, here the $\Ro^3$ spanned by the
velocities $\mathbf v$,  is time independent. The time derivative
and the derivative with respect to position can be interchanged
with the integration over $\mathcal M$. We split the mesoscopic
velocity  into the macroscopic velocity  and the deviation from
this average, respectively, for component A and  the mixture: \be
\mathbf v(\cdot)=\mathbf{v}^A(\mathbf x,t) +\delta\mathbf v^A ,
\ee and \be \mathbf v(\cdot)=\mathbf{v}(\mathbf x,t)
+\delta\mathbf v. \ee The resulting equations of motion for the
first moments $\mathbf a^A_1$ and $\mathbf a_1$ are, respectively:
\beg \label{eq:fmA} &&\frac{\partial}{\partial t}\mathbf
a^A_1+\nabla\cdot\Big[\mathbf{v}^A (\mathbf x,t)\mathbf a^A_1
+\int_{\Ro^3}\delta\mathbf{v}^A(\cdot) f^A(\cdot)\mathbf v^A d\mathcal M\Big]+\\
\nonumber &&\int_{\Ro^3}\mathbf v^A\nabla_{\mathbf
v}\cdot[f^A(\cdot)\mathbf w^A(\cdot)] d\mathcal M+\mathbf a^A_1
\Big[\frac{\partial}{\partial t} +\mathbf{v}^A(\mathbf x,t)\cdot
\nabla \Big]\lg\rho^A(\mathbf x,t)+\\ \nonumber &&+\nabla
\lg\rho^A(\mathbf x,t) \int_{\Ro^3}\delta\mathbf{v}_A (\cdot)
f^A(\cdot)\mathbf v^A d\mathcal M= \frac 1{\rho^A(\mathbf
x,t)}\int\hat P^A_{chem}\mathbf v^A d\mathcal M \ene and
\beg\nonumber &&\frac{\partial}{\partial t}\mathbf
a_1+\nabla\cdot\Big[\mathbf{v}(\mathbf x,t)\mathbf a_1
+\int_{\Ro^3}(\delta\mathbf{v}(\cdot) f(\cdot)\mathbf v d\mathcal
M\Big]+\\                     \label{eq:fm} &&\int_{\Ro^3}\mathbf
v\nabla_{\mathbf v}\cdot[f(\cdot)\mathbf w(\cdot)] d\mathcal
M+\mathbf a_1 \Big[\frac{\partial}{\partial t}+ \mathbf{v}(\mathbf
x,t)\cdot \nabla \Big]\lg\rho(\mathbf x,t)+\\ \nonumber &&+\nabla
\lg\rho(\mathbf x,t) \int_{\Ro^3}\delta\mathbf{v} (\cdot)
f(\cdot)\mathbf v d\mathcal M=0 \ene We now introduce some
approximations: the deviations  $\delta \mathbf v^A$ and $\delta
\mathbf v$ vanish, i.e., the mesoscopic variable $\mathbf v$ has
the value of the barycentric velocity. In other words we suppose
$\mathbf{v}^A(\cdot)=\mathbf{v}^A (\mathbf x,t)$ and
$\mathbf{v}(\cdot)=\mathbf{v}(\mathbf x,t)$.\newline By computing
the difference of equations (\ref{eq:fmA}) and (\ref{eq:fm}) and
taking into account these approximations one has: \beg\nonumber
&&\frac{\partial}{\partial t}(\mathbf a^A_1-\mathbf a_1)+
\nabla\cdot (\mathbf{v}^A\mathbf a^A_1-\mathbf{v} \mathbf a_1)+
\int_{\Ro^3} \mathbf v \nabla_{\mathbf v}\cdot (f^A(\cdot)\mathbf
w^A(\cdot)\\\nonumber &&- f(\cdot)\mathbf w(\cdot)d\mathcal
M+\mathbf a^A_1 \Big[\frac{\partial}{\partial t} +\mathbf{v}^A(\mathbf x,t)\cdot \nabla \Big] \lg\rho^A(\mathbf x,t)\\
&&-\mathbf a_1 \Big[\frac{\partial}{\partial t}
+\mathbf{v}(\mathbf x,t)\cdot \nabla \Big]\lg\rho(\mathbf
x,t)\label{eq:evflux}=\frac 1{\rho(\mathbf x,t)}\int_{\Ro^3}\hat
P^A_{chem}d\mathcal M \ene The last equation (\ref{eq:evflux})
represents a general phenomenological evolution equation for
diffusion fluxes in an "extended theory". Using further
approximations   it is possible to obtain a Cattaneo-type
diffusion equation. In order to exploit equation (\ref{eq:evflux})
further, it is necessary to insert expressions for $\mathbf w$ and
$\mathbf w^A$. These are equations on the mesoscopic level, and
they can be interpreted  as mesoscopic constitutive equations. In
this case the domain of constitutive mappings, the state space has
to be introduced. In mesoscopic theories there are different
possibilities: one is to introduce a state space of purely
mesoscopic variables, another one is to introduce a mixed space
including mesoscopic and macroscopic variables, and the third
possibility is one containing only macroscopic variables. The
mesoscopic background of the constitutive theory has been applied
to liquid crystals of uniaxial molecules
\cite{PhysRev95,Bam,max_ent,Arch} and to dipolar media
\cite{PaCianRo02}. For diffusion in mixtures this is left for
future work.

\section{Conclusions}

We have shown a way to derive a differential equation for the
diffusion flux from the so called mesoscopic theory. In this
refined theory an additional variable in the domain of the field
quantities is introduced, here the  velocity, and we have a
velocity distribution. We have derived a differential equation for
the diffusion flux from the mesocopic balance equations. The
differential equation for the diffusion flux is of balance type,
i.e., of the form used in extended thermodynamics. Therefore, it
can be expected that our approach leads to hyperbolic systems of
field equations, but the investigation of this question is left
for future work.

Because the level here is not the microscopic one, no assumptions
about inter-particle interactions have to be made here in contrast
to kinetic theory. We are here on the continuum level, and
many-particle interactions are taken into account automatically.
This is not the case for derivations of the field equations of
extended thermodynamics based on the Boltzmann-equation.

\section{Acknowledgements}

We are grateful to W. Muschik for stimulating discussions and
remarks. This work was supported by Fondi P.R.A 2002, University
of Messina.


\begin{thebibliography}{10}

\bibitem{Jou}
D.~Jou, J.~Casas-Vazquez, and G.~Lebon.
\newblock {\em Extended Irreversible Thermodynamics}.
\newblock Springer-Verlag, Berlin, Heidelberg, New York, 1993.

\bibitem{Jou92}
D.~Jou, J.~Casas-Vazquez, and G.~Lebon.
\newblock Extended irreversible thermodynamics: an overview of recent
  bibliography.
\newblock {\em J. Non-Equilb. Thermodyn.}, 17:383--396, 1992.

\bibitem{Jou96}
D.~Jou, J.~Casas-Vazquez, and G.~Lebon.
\newblock Recent bibliography on extended irreversible thermodynamics and
  related topics (1992-1995).
\newblock {\em J. Non-Equilb. Thermodyn.}, 21:103--121, 1996.

\bibitem{15Jou}
D.~Jou, J.~Casas-Vasquez, and G.~Lebon.
\newblock Extended irreversible thermodynamics.
\newblock {\em Rep. Prog. Phys.}, 51:1105--1179, 1988.

\bibitem{MuellerRuggeri}
I.~Mueller and T.~Ruggeri.
\newblock {\em Extended Thermodynamics}, volume~37.
\newblock Springer Tracts in Natural Philosophy, Berlin, Heidelber, New York,
  1993.


\bibitem{Lebon01}
N. ~Depiteux, G. ~Lebon.
\newblock An extended thermodynamics modelling
of non-Fickian diffusion,
\newblock {\em J. Non-Newtonian Fluid Mech.} 96, 2001.



\bibitem{Garcia_94}
P.~Goldstein and L.~S. Garcia-Colin.
\newblock Transport processes in a viscoelastic binary mixture.
\newblock {\em J. Non-Equilib. Thermodyn.}, 19:170--183, 1994.

\bibitem{Garcia_93}
P.~Goldstein and L.~S. Garcia-Colin.
\newblock A thermodynamic basis for transport phenomena in viscoelastic fluids.
\newblock {\em J. Chem. Phys.}, 99:3913, 1993.

\bibitem{Jou_1}
D.~Jou, J.~Casas-Vazquez, and M.~Criado-Sancho.
\newblock Polymer solutions under flow: phase separation and polymer
  degradation.
\newblock {\em Advances in {P}olymer {S}cience}, 120:205--266, 1995.

\bibitem{Lebon_94}
G.~Lebon, D.~Jou, and J.~Casas-Vazquez.
\newblock Nonequilibrium entropy and the second law of thermodynamics: a simple
  illustration.
\newblock {\em Int. J. Thermophys}, 14:671--683, 1993.

\bibitem{Casas_93}
J.~Casas-Vazquez, M.~Criado-Sancho, and D.~Jou.
\newblock Dynamical and thermodynamical approaches to phase separation in
  polymer solutions under flow.
\newblock {\em Europhys. Lett.}, 23:469--474, 1993.

\bibitem{Criado_92}
M.~Criado-Sancho, D.~Jou, and J.~Casas-Vazquez.
\newblock On the spinodal line of polymer solutions under shear.
\newblock {\em J. Non-Equilib. Thermodyn.}, 18:103--120, 1992.

\bibitem{Criado_93}
G.~Lebon, J.~Casas-Vazquez, M.~Criado-Sancho, and D.~Jou.
\newblock Polymer solutions and chemical reactions under flow: a thermodynamic
  description.
\newblock {\em J. Chem. Phys.}, 98:7434--7439, 1993.

\bibitem{MULLER}
I.~Mueller.
\newblock {\em Thermodynamics}.
\newblock Pitman Advanced Publishing Program, Boston, London, Melbourne, 1985.

\bibitem{Mueller_Rev}
I.~Mueller.
\newblock Extended thermodynamics of classical and degenerate gases.
\newblock {\em Arch. Rat. Mech. Anal.}, 83:286--332, 1983.

\bibitem{Van_Cimelli_04}
V.~A.~Cimelli, and P.~V\'{a}n.
\newblock The effects of non-locality on the evolution of  higher order fluxes in non-equilibrium
thermodynamics.
\newblock cond-mat/0409254, 2004.


\bibitem{Van_Cimelli_05}
V.~Ciancio, V.~A.~Cimelli, and P.~V\'{a}n.
\newblock Balance laws for higher order fluxes in non-equilibrium
thermodynamics.
\newblock cond-mat/0407530, 2005.

\bibitem{Ruggeri83}
T.~Ruggeri.
\newblock Symmetric-hyperbolic system of conservative equations for a viscous
  heat conducting fluid.
\newblock {\em Acta Mechanica}, 77(3):167--183, 1983.
\newblock Review Article.

\bibitem{Weiss_dr}
W.~Weiss.
\newblock {\em Zur Hierarchie der Erweiterten Thermodynamik. Thesis}.
\newblock TU Berlin, Berlin, 1990.

\bibitem{Friedrichs_Lax}
K.~O. Friedrichs and P.~D. Lax.
\newblock Systems of conservation equations with a convex extension.
\newblock {\em Proc. Nat. Acad. Sci USA}, 68:1686--1688, 1971.

\bibitem{Lax73}
P.~D. Lax.
\newblock {\em Hyperbolic systems of conservation laws and the mathematical
  theory of shock waves}.
\newblock Society for Industrial and Applied Mathematics, Bristol, 1973.

\bibitem{BornGreen46}
M.~Born and H.~S. Green.
\newblock {\em Proc. Roy. Soc. Lond. A}, 188:10, 1946.

\bibitem{BornGreen47}
M.~Born and H.~S. Green.
\newblock {\em Proc. Roy. Soc. Lond. A}, 190:455, 1947.

\bibitem{Bogoljuboff}
N~Bogoljuboff.
\newblock {\em J. Phys. USSR}, 10:265, 1946.

\bibitem{Kirkwood}
J.~G. Kirkwood.
\newblock {\em J. Chem. Phys.}, 14:180, 1946.

\bibitem{Sitges}
W.~Muschik, H.~Ehrentraut, C.~Papenfuss, and S.~Blenk.
\newblock Mesoscopic theory of liquid crystals.
\newblock In {\em J.J.Brey, J.Marro, J.M.Rubi, M.San Miguel (Eds.): 25 Years of
  Non-Equilibrium Statistical Mechanics, Proceedings of the XIII Sitges
  Conference, 13 - 17 June 1994, Sitges}, volume 445 of {\em Lecture Notes in
  Physics}, pages 303--311. Springer, 1995.
\bibitem{MUPAEH00}
W.~Muschik, C. ~Papenfuss, H. ~Ehrentraut.
\newblock Concept of mesoscopic
continuum physics with application to biaxial Liquid Crystals.
\newblock {\em J. Non-Equilib. Thermodyn.}, 26, 2000.
\bibitem{PACIRO02}
C. ~Papenfuss, V. ~Ciancio, P. ~Rogolino.
\newblock Application of the mesoscopic theory to
dipolar media,
\newblock{\em Technische Mechanik}, 22 (2), 2002.


\bibitem{MUPAEH96}
~Muschik, W, C.~Papenfuss, and H.~Ehrentraut.
\newblock {\em Concepts of Continuum Thermodynamics}.
\newblock Kielce University of Technology, Technische Universit\"at Berlin,
  ISBN 83-905132-7-7, 1996.

\bibitem{poz}
C.~Papenfuss and W.~Muschik.
\newblock Liquid crystal theory as an example of mesoscopic continuum
  mechanics.
\newblock In B.~T. Maruszewski, W.~Muschik, and A.~Radowicz, editors, {\em
  Trends in Continuum Physics}, pages 277--291. World Scientific, Singapore,
  1998.

\bibitem{BLENK91}
S.~Blenk and W.~Muschik.
\newblock Orientational balances for nematic liquid crystals.
\newblock {\em J. Non-Equilib. Thermodyn.}, 16:67--87, 1991.

\bibitem{BLENK92}
S.~Blenk, H.~Ehrentraut, and W.~Muschik.
\newblock Macroscopic constitutive equations for liquid crystals induced by
  their mesoscopic orientation distribution.
\newblock {\em Int. J. Engng. Sci.}, 30(9):1127--1143, 1992.

\bibitem{PHYSICA}
S.~Blenk, H.~Ehrentraut, and W.~Muschik.
\newblock Statistical foundation of macroscopic balances for liquid crystals in
  alignment tensor formulation.
\newblock {\em Physica A}, 174:119--138, 1991.

\bibitem{PhysRev95}
Harald Ehrentraut and Siegfried Hess.
\newblock Viscosity coefficients of partially aligned nematic and nematic
  discotic liquid crystals.
\newblock {\em Phys. Rev. E}, 51(3):2203 -- 2212, March 1995.

\bibitem{Journal}
H.~Ehrentraut, W.~Muschik, and C.~Papenfuss.
\newblock Mesoscopically derived orientation dynamics of liquid crystals.
\newblock {\em J. Non-Equilib. Thermodyn.}, 22:285--298, 1997.

\bibitem{Budapest2}
W.~Muschik, H.~Ehrentraut, and C.~Papenfuss.
\newblock The connection between ericksen-leslie equations and the balances of
  mesoscopic theory of liquid crystals.
\newblock {\em Mol. Cryst. Liq. Cryst.}, 262:417--423, 1995.

\bibitem{mesocrack}
P.~Van, C.~Papenfuss, and W.~Muschik.
\newblock Mesoscopic dynamics of microcracks.
\newblock {\em Physical Review E}, 62(5):6206--6215, 2000.

\bibitem{VanPa02}
P.~V\'an, C.~Papenfuss, and W.~Muschik.
\newblock Griffith cracks in the mesoscopic microcrack theory.
\newblock published online: Condensed Matter, abstract, cond-mat/0211207; sent
  to Phy. Rev. E, 2002.

\bibitem{Ehrentraut_diss}
H.~Ehrentraut.
\newblock {\em A {U}nified {M}esoscopic {C}ontinuum {T}heory of {U}niaxial and
  {B}iaxial {L}iquid {C}rystals}.
\newblock Wissenschaft und Technik Verlag, Berlin, 1996.

\bibitem{MCLC}
S.~Blenk, H.~Ehrentraut, and W.~Muschik.
\newblock Orientation balances for liquid crystals and their representation by
  alignment tensors.
\newblock {\em Mol. Cryst. Liqu. Cryst.}, 204:133--141, 1991.

\bibitem{Bam}
C.~Papenfuss.
\newblock Theory of liquid crystals as an example of mesoscopic continuum
  mechanics.
\newblock {\em Computational Materials Science}, 19:45 -- 52, 2000.

\bibitem{MUSCHIK83a1}
W.H. M\"uller and W.~Muschik.
\newblock {B}ilanzgleichungen offener mehrkomponentiger {S}ysteme \\ {I}.
  {M}assen- und {I}mpulsbilanzen.
\newblock {\em J. Non-Equilib. Thermodyn.}, 8:29--46, 1983.

\bibitem{MUSCHIK83a2}
W.~Muschik and W.H. M\"uller.
\newblock {B}ilanzgleichungen offener mehrkomponentiger {S}ysteme \\ {II}.
  {E}nergie- und {E}ntropiebilanz.
\newblock {\em J. Non-Equilib. Thermodyn.}, 8:47--66, 1983.

\bibitem{Blenk93}
S.~Blenk, H.~Ehrentraut, and W.~Muschik.
\newblock A continuum theory for liquid crystals describing different degrees
  of orientational order.
\newblock {\em Liquid Crystals}, 14(4):1221--1226, 1993.



\bibitem{max_ent}
C.~Papenfuss and W.~Muschik.
\newblock Orientational order in free standing liquid crystalline films and
  derivation of a closure relation for higher order alignment tensors.
\newblock {\em Mol. Cryst. Liq. Cryst.}, 330:541 -- 548, 1999.

\bibitem{Arch}
C.~Papenfuss.
\newblock Nonlinear dynamics of the alignment tensor in the presence of
  electric fields.
\newblock {\em Arch. Mech.}, 50(3):529--536, 1998.



\bibitem {PaCianRo02}
  C.~Papenfuss, V.~Ciancio, and P.~Rogolino. \newblock
  Application of the mesoscopic theory to dipolar media.
  \newblock     { \em Technische Mechanik}
  22  (2) (2002).




\end{thebibliography}


\end{document}